\documentclass[sigconf]{acmart}

\usepackage{booktabs}
\usepackage{amssymb}
\usepackage{amsmath} 
\usepackage{courier}
\usepackage{enumitem}
\usepackage{overpic}
\usepackage{graphicx}
\usepackage{subfigure}
\usepackage{bbding}
\usepackage{pifont}
\usepackage{setspace}
\usepackage{array,multirow,multicol}
\usepackage{caption, subcaption}
\usepackage{algorithm} 
\usepackage{makecell}
\usepackage{algpseudocode} 

\usepackage[frozencache, cachedir=minted-cache]{minted}
\usepackage{xcolor}
\usepackage{tcolorbox}
\tcbuselibrary{skins}
\tcbuselibrary{minted}
\usepackage{url}

\newcommand{\PreserveBackslash}[1]{\let\temp=\\#1\let\\=\temp}
\newcolumntype{C}[1]{>{\PreserveBackslash\centering}p{#1}}
\newcolumntype{R}[1]{>{\PreserveBackslash\raggedleft}p{#1}}
\newcolumntype{L}[1]{>{\PreserveBackslash\raggedright}p{#1}}
  
\usepackage[normalem]{ulem} 

\AtBeginDocument{%
}

\copyrightyear{2024}
\acmYear{2024}
\setcopyright{rightsretained}
\acmConference[SIGIR '24]{Proceedings of the 47th International ACM SIGIR Conference on Research and Development in Information Retrieval}{July 14--18, 2024}{Washington, DC, USA}
\acmBooktitle{Proceedings of the 47th International ACM SIGIR Conference on Research and Development in Information Retrieval (SIGIR '24), July 14--18, 2024, Washington, DC, USA}
\acmDOI{10.1145/3626772.3657868}
\acmISBN{979-8-4007-0431-4/24/07}

\begin{document}
\begin{sloppypar}

\title{EasyRL4Rec: An Easy-to-use Library for Reinforcement Learning Based Recommender Systems}


\author{Yuanqing Yu}
\affiliation{
  \institution{DCST, Tsinghua University}
  \city{Beijing 100084}
  \country{China}
  \postcode{100084}
}
\email{yyq23@mails.tsinghua.edu.cn}

\author{Chongming Gao}
\authornote{Corresponding author. This work is supported by the Natural Science Foundation of China (Grant No. U21B2026, 62372399, 62372260), and the advanced computing resources provided by the Supercomputing Center of Hangzhou City University.}
\affiliation{
  \institution{University of Science and Technology of China}
  \country{}
}
\email{chongming.gao@gmail.com}

\author{Jiawei Chen}
\authornotemark[1]
\affiliation{%
  \institution{Zhejiang University}
  \city{Hangzhou}
  \country{China}
}
\email{sleepyhunt@zju.edu.cn}

\author{Heng Tang}
\affiliation{
  \institution{Zhejiang University}
  \city{Hangzhou}
  \country{China}
}
\email{tanghengsboy@gmail.com}

\author{Yuefeng Sun}
\affiliation{
  \institution{Zhejiang University}
  \city{Hangzhou}
  \country{China}
}
\email{sun920523033@163.com}

\author{Qian Chen}
\affiliation{
  \institution{University of Science and Technology of China}
  \country{}
}
\email{chenxing202022@gmail.com	}

\author{Weizhi Ma}
\affiliation{%
  \institution{AIR, Tsinghua University}
  \city{Beijing 100084}
  \country{China}
  \postcode{100084}
}
\email{mawz@tsinghua.edu.cn}

\author{Min Zhang}
\affiliation{%
  \institution{DCST, Tsinghua University}
  \city{Beijing 100084}
  \country{China}
  \postcode{100084}
}
\email{z-m@tsinghua.edu.cn}

\renewcommand{\shortauthors}{Yuanqing Yu, et al.}

\begin{abstract}
Reinforcement Learning (RL)-Based Recommender Systems (RSs) have gained rising attention for their potential to enhance long-term user engagement. However, research in this field faces challenges, including the lack of user-friendly frameworks, inconsistent evaluation metrics, and difficulties in reproducing existing studies. To tackle these issues, we introduce \textit{EasyRL4Rec}, an easy-to-use code library designed specifically for RL-based RSs. 
This library provides lightweight and diverse RL environments based on five public datasets and includes core modules with rich options, simplifying model development.
It provides unified evaluation standards focusing on long-term outcomes and offers tailored designs for state modeling and action representation for recommendation scenarios. 
Furthermore, we share our findings from insightful experiments with current methods.
EasyRL4Rec seeks to facilitate the model development and experimental process in the domain of RL-based RSs.
The library is available for public use\footnote{\url{https://github.com/chongminggao/EasyRL4Rec}}.
\end{abstract}



\maketitle

\section{Introduction}
Recommender systems (RSs) are increasingly becoming integral to various domains, such as e-commerce, social media, and online streaming services. Traditional RSs usually rely on supervised learning methods like collaborative filtering~\cite{schafer2007collaborative} to learn and predict users' interests. However, these methods often fail to capture the long-term effects, leading to potential issues like Feedback Loops~\cite{mansoury2020feedback}, Filter Bubbles~\cite{flaxman2016filterbubbles}, and other undesirable biases~\cite{chen2023bias, chen2021autodebias}.

Reinforcement Learning (RL)-Based Recommender Systems have gained rising attention due to their ability to optimize long-term user engagement.
In this setting, the recommendation process is perceived as a multi-step decision-making problem. An agent (recommender) interacts with the environment (users) and receives feedback on actions (items). The main goal of RL-based RSs is to learn an optimal recommendation policy that maximizes cumulative rewards (users' long-term engagement) through trial and error. 
Currently, RL-based RSs have been extensively applied in scenarios such as short video recommendations~\cite{Kuai-cai2023two, Kuai-liu2023exploration, Kuai-cai2023rein, Kuai-xue2023-prefrec}, e-commerce recommendations~\cite{rohde2018recogym, zhao2018deep}, and beyond.

\begin{table*}[t!]
\small
\centering
\caption{A comparison between EasyRL4Rec and existing resources. Real Data refers to using data collected from the real world and not generated by simulators. Disc.Act. indicates support for discrete action-based policies, while Cont.Act. is short for continuous action. Interactive training indicates learning with instant feedback.}
\label{tab:existing_resources}
\begin{tabular}{cccccccccc}
\toprule
\multirow{2}{*}{\textbf{Type}} & \multirow{2}{*}{\textbf{Resource}} & \multicolumn{5}{c}{\textbf{Modules}} & \multicolumn{2}{c}{\textbf{Training}}  & \textbf{Evaluation} \\
\cmidrule(lr){3-7}\cmidrule(lr){8-9}\cmidrule(lr){10-10}
 &  & \textbf{Real Data} & \textbf{State Encoder} & \textbf{Disc.Act.} & \textbf{Cont.Act.} & \textbf{RL Policy} & \textbf{Offline Logs} & \textbf{Interactive} & \textbf{Long-term} \\
 

\midrule
\multirow{6}{*}{\begin{tabular}[c]{@{}c@{}}Simulators\\ \& Datasets\end{tabular}} 
 & RecoGym~\cite{rohde2018recogym} &  &  & \CheckmarkBold &  &  &  \CheckmarkBold & \CheckmarkBold &  \\
 & RecSim~\cite{ie2019recsim} &  &  & \CheckmarkBold &  & \CheckmarkBold &  & \CheckmarkBold  & \CheckmarkBold \\
 & Virtual-Taobao~\cite{shi2019virtual} & \CheckmarkBold &  & \CheckmarkBold & \CheckmarkBold & \CheckmarkBold &   & \CheckmarkBold & \CheckmarkBold \\
 & SOFA~\cite{huang2020sofa} & \CheckmarkBold & \CheckmarkBold & \CheckmarkBold &  & \CheckmarkBold & \CheckmarkBold &  & \CheckmarkBold \\
 & KuaiSim~\cite{zhao2023kuaisim}& \CheckmarkBold &  & \CheckmarkBold &  & \CheckmarkBold &  & \CheckmarkBold & \CheckmarkBold \\
 & RL4RS~\cite{wang2023rl4rs} & \CheckmarkBold &  & \CheckmarkBold & \CheckmarkBold &  \CheckmarkBold & \CheckmarkBold & \CheckmarkBold & \CheckmarkBold \\
\midrule
\multirow{5}{*}{\begin{tabular}[c]{@{}c@{}}Frameworks\\ \& Libraries\end{tabular}} 
 & BEARS~\cite{barraza2018bears} & \CheckmarkBold &  & \CheckmarkBold &  & \multirow{4}{*}{\begin{tabular}[c]{@{}c@{}}Only have\\Bandits\\methods.\end{tabular}} & \multicolumn{2}{c}{\multirow{4}{*}{\begin{tabular}[c]{@{}c@{}}Most Bandits methods \\do not have an  \\explicit training.\end{tabular}}} & \CheckmarkBold \\
 & MABWiser~\cite{strong2019mabwiser} & \CheckmarkBold &  & \CheckmarkBold &  &  &  &  & \CheckmarkBold \\
 & OBP~\cite{saito2020obp} & \CheckmarkBold &  & \CheckmarkBold &  &  &  &  & \CheckmarkBold \\
 & iRec~\cite{silva2022irec} & \CheckmarkBold &  & \CheckmarkBold &  &  &  &  & \\
\cmidrule(lr){2-10}
  & \textbf{EasyRL4Rec} & \CheckmarkBold & \CheckmarkBold & \CheckmarkBold & \CheckmarkBold & \CheckmarkBold & \CheckmarkBold & \CheckmarkBold & \CheckmarkBold \\
\bottomrule
\end{tabular}
\end{table*}

With increasing attention to RL-based RSs, research in this field is confronted with the following three challenges.
\begin{itemize}[leftmargin=*, topsep=0pt, parsep=0pt]
\item \textbf{Absence of an easy-to-use framework.} Currently, there is a lack of easy-to-use frameworks for academic research. Research for RL-based RSs requires interactive environment building. However, existing resources neither use mass datasets nor simulated data generated by simulators, making them inconvenient for usage.
In addition, it is difficult to directly apply RL libraries (e.g. rllib~\cite{liang2018rllib}) to recommendation scenarios due to the absence of suitable environments and state modeling.
\item \textbf{Distinct evaluation strategies in different work.}
The absence of standardized evaluation metrics complicates model comparisons across different research teams. 
While some studies follow traditional RS metrics such as Normalized Discounted Cumulative Gain (NDCG) and Hit Rate (HR), others use RL-specific measures like cumulative reward and interaction length, leading to inconsistencies in performance assessment.
\item \textbf{Poor reproducibility of previous studies.}
Thirdly, poor reproducibility of some previous studies increases the difficulty of further studies. 
Researchers and practitioners in this field often face challenges in reproducing methods and ensuring the effectiveness of self-implemented baselines due to variations in details, hindering the field's development.
\end{itemize}

To tackle the above issues, we implement a comprehensive code library for RL-based RSs, named \textit{EasyRL4Rec}. It provides an easy-to-use framework with core modules with rich choices and a unified training and evaluating process, aiming to simplify the model development and experimental process in the domain of RL-based RS.
The library is composed of four core modules: \texttt{Environment}, \texttt{Policy}, \texttt{StateTracker}, and \texttt{Collector}, which cater to different stages of the RL interaction process. 
\texttt{Environment}, built from lightweight static datasets, provides feedback on upcoming actions. The \texttt{Policy} module selects the optimal action based on the current state, which is encoded by \texttt{StateTracker}. \texttt{Collector} bridges the interactions between \texttt{Environment} and \texttt{Policy}. 
To address the difficulty of obtaining user states from environments, we implement multiple StateTrackers for state modeling, encompassing today's popular methods in sequential modeling~\cite{GRU4Rec-hidasi2015session, Caser-tang2018personalized, SASRec-kang2018self, NextItNet-yuan2019simple}. Since items in RSs are discrete, EasyRL4Rec includes a mechanism to convert continuous actions to discrete items, allowing for continuous action-based policies.

Moreover, EasyRL4Rec provides a unified training and evaluation procedure by \texttt{Trainer} and \texttt{Evaluator} executors. \texttt{Evaluator} supports three modes, allowing for the removal of recommended items and a quit mechanism.  \texttt{Trainer} offers two training paradigms: learning directly from offline logs or training with a pre-trained user model. 
With this unified framework, we conduct comprehensive experiments on classic RL models and several recent work~\cite{Self-Supervised-RLinRec-Xinxin, DORL-gao2023alleviating} and present insightful results.
The main contributions of this work can be summarized as follows:
\begin{itemize}[leftmargin=*] 
    \item \textbf{Easy-to-use Framework}. 
    EasyRL4Rec provides an easy-to-use framework for RL-based RSs.
    We construct lightweight RL environments based on five public datasets encompassing diverse domains, which are easy to follow for researchers.
    Moreover, the design of core modules with rich options reduces the complexity of developing a new model.
    \item \textbf{Unified Evaluation Standards}. 
    EasyRL4Rec offers a unified experimental pipeline, evaluating models with various metrics from the perspective of long-term benefits (e.g. Cumulative Reward). Furthermore, the library offers two training paradigms and three evaluation settings, giving users multiple choices. 
    
    \item \textbf{Tailored Designs for Recommendation Scenarios}.
    In response to challenges when applying RL algorithms in practical RSs, we have developed customizable modules for state modeling and action representation, with a conversion mechanism to support continuous action-based policies.
    
    \item \textbf{Insightful Experiments for RL-based RSs}. With EasyRL4Rec, we conduct experiments to compare the performance of classic RL models and some recent work, presenting insightful results from multiple perspectives.
\end{itemize}

\section{Related work}

\subsection{Development of RL-Based RSs}

Reinforcement Learning (RL), a branch of machine learning, focuses on agents learning the ability of decision-making through environmental feedback\footnote{RL in this work specifically refers to deep reinforcement learning.}. 
Recently, RL-Based Recommender Systems have gained considerable attention due to their ability to model recommendation as a multi-step decision-making process and enhance the long-term benefits.

Numerous studies have investigated the application of RL in recommender systems.
~\citet{shani2005mdp} first formulated the recommendation process as a Markov Decision Process (MDP) and utilized model-based RL methods.
\citet{DEERS-RLinRec-zhao2018recommendations} adapt a DQN~\cite{DQN-mnih2013playing} architecture to incorporate positive and negative feedback from users. ~\citet{DRN-RLinRec-zheng2018drn} applies the Dueling DQN algorithm to news recommendation.  \citet{REINFORCE-RLinRec-chen2019top, offpolicy-chen2022off} extend REINFORCE~\cite{REINFORCE-williams1992simple} and off-policy actor-critic algorithm to recommendation.
~\citet{Self-Supervised-RLinRec-Xinxin} proposed to utilize self-supervision signals to empower RL-based RSs.
Unlike most methods based on discrete actions, ~\citet{Kuai-cai2023rein, Kuai-liu2023exploration, Kuai-xue2023-prefrec} investigate the applications of continuous action-based RL and improvements of long-term user engagement in short-video scenarios.
In addition, ~\citet{ren2023contrastive} focus on effective state representations, while recent work~\cite{CIRS-gao2022cirs, DORL-gao2023alleviating} addresses specific issues when applying RL in RSs.

The above work mainly focuses on optimizing RL policy, state modeling, or specific issues in recommendation scenarios. However, over 60\% of existing work is not open-source, inspiring us to develop a library that encompasses these aspects and facilitates the reproduction of existing work.



\subsection{Resources for RL-Based RSs}

Existing resources for RL-based RSs can be categorized into two groups: Simulators \& Datasets, and Frameworks \& Libraries.
In Table ~\ref{tab:existing_resources}, we summarize the characteristics of existing resources regarding modules and experimental procedures.
\subsubsection{Simulators and Datasets}
Numerous studies have focused on the development of datasets or simulation platforms to create interactive RL environments.
RecoGym~\cite{rohde2018recogym} focuses on simulation bandit environments under e-commerce recommendation setting, while Virtual-Taobao~\cite{shi2019virtual} provides a user simulator trained on historical behavior data.
RecSim~\cite{ie2019recsim} is a simulation platform supporting sequential user interactions and easy configuration of environments. Additionally, SOFA~\cite{huang2020sofa} is the first simulator that accounts for interaction biases for optimization and evaluation.
In the latest work, RL4RS~\cite{wang2023rl4rs} provides a validated simulation environment, advanced evaluation methods, and a real dataset. KuaiSim~\cite{zhao2023kuaisim} offers an environment with multi-behavior feedback, supporting three levels of recommendation tasks.

However, most of these studies merely provide an interactive environment, lacking integrated experimental procedures and other crucial components in RL-based RSs, such as diverse policies and state trackers. 
RL4RS is the most similar work to our library, with the following distinctions: 1) They focus on datasets and evaluation, whereas our goal is to offer an easy-to-use framework for developing new models and facilitating experiments. 
2) They support only specific slate recommendation scenarios, whereas our library caters to broader and more commonly used scenarios.

\subsubsection{Frameworks and Libraries}

As far as we know, there are few frameworks or libraries that directly address RL-based RSs. However, some high-quality libraries exist for Multi-Armed Bandit (MAB) algorithms~\cite{UCB-auer2002finite, Egreedy-tokic2010adaptive, ThompsonSampling-chapelle2011empirical, LinUCB-li2010contextual} with interactive training environments\footnote{In Table~\ref{tab:existing_resources}, MAB algorithms is not classified as RL policies, within the scope of deep reinforcement learning.}.
BEARS~\cite{barraza2018bears} serves as an evaluation framework that facilitates easy testing of bandit-based RS solutions and supports reproducible offline assessments. MABWiser~\cite{strong2019mabwiser} is a parallelizable library that supports traditional MAB solutions and Contextual Bandits. Open Bandit Pipeline (OBP)~\cite{saito2020obp} focuses on off-policy evaluation (OPE) and provides a streamlined and standardized library for implementing batch bandit algorithms and OPE. iRec~\cite{silva2022irec} proposes an interactive recommender systems framework that also caters to multi-armed bandits models.

These libraries primarily serve bandit-based RSs and do not incorporate RL policies. Moreover, some well-established libraries designed for classic RL scenarios(e.g. rllib~\cite{liang2018rllib}, tianshou~\cite{tianshou}) cannot be directly applied to recommendations due to the lack of recommendation environment construction and state modeling.





\section{Preliminaries}


In RL-based RSs, the sequential interactions are formulated as a Markov decision process (MDP) $M=(\mathcal S,\mathcal A,\mathcal P,R,\gamma)$ where
\begin{itemize}[itemsep=0pt,topsep=0pt,parsep=0pt]
    \item $\mathcal S$, state space, $s_t \in \mathcal S$ represents the current state of user at timestamp $t$. In recommendation scenarios, user characteristics and user history are usually modeled as states, where user history refers to the actions and corresponding feedback of each step in the previous interaction process.
    \item $\mathcal A$, action space, $a_t \in \mathcal A$ represents the action that RL agent take at timestamp $t$. In recommendation scenarios, the recommended product is generally chosen as the action.
    \item $\mathcal P$: state transition probability function, $P(s,a,s^{\prime})=P(s_{t+1}=s^{\prime}|s_t=s,a_t=a)$ represents the transition probability from $(s,a)$ to $s^{\prime}$.
    \item $r$: reward function, $r(s,a)$ denotes the reward by taking action $a$ at state $s$. Rewards in recommendation scenarios are usually set to feedback signals provided by users, such as click behavior, favorite behavior, or dwell time.
    \item $\gamma$: the discount factor for future rewards.
\end{itemize}

The main goal of RL is to learn an optimal decision-making policy $\uppi_\theta(a|s)$to maximize the cumulative reward $G_T$: 
\begin{equation}
    \label{eq:max_cumulative_reward}
    \max_{\uppi}\mathbb{E}[G_T] = \max_{\uppi}\mathbb{E}\Big[\sum_{t=0}^T \gamma^t r(s_t,a_t)\Big]
\end{equation}


\begin{figure}[t!]
\centering
\includegraphics[trim={0 0 0 0}, clip, width=0.46\textwidth]{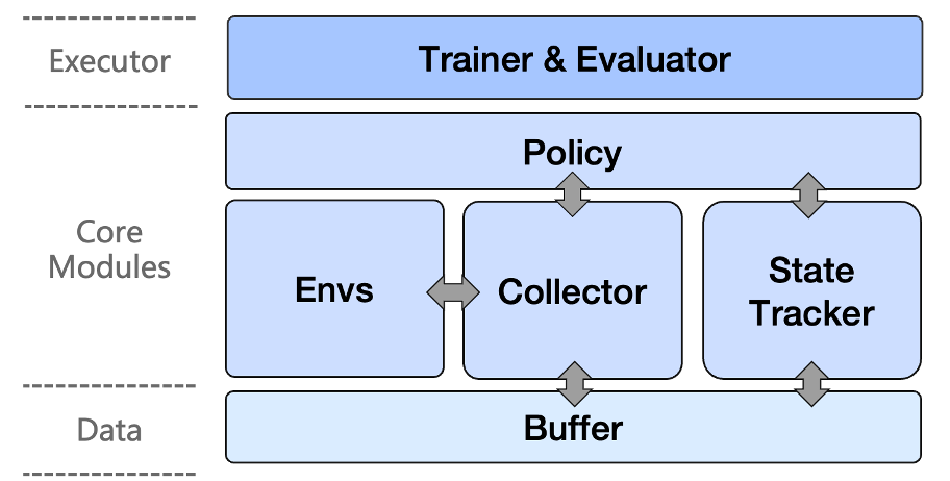}
\caption{Architecture of EasyRL4Rec. The library is structured around four core modules: \texttt{Environment} (abbreviated as Envs), \texttt{Policy}, \texttt{StateTracker}, and \texttt{Collector}. \texttt{Buffer} serves as a fundamental data structure for organizing raw data trajectories, while the \texttt{Trainer} and \texttt{Evaluator} act as executors, managing the entire process.}
\label{fig:architecture}
\end{figure}

\section{The Library-EasyRL4Rec}

\subsection{Library Overview}

The overall architecture of EasyRL4Rec is illustrated in Figure~\ref{fig:architecture}. The library is composed of four core modules: \texttt{Environment}, \texttt{Policy}, \texttt{StateTracker}, and \texttt{Collector}, each addressing distinct stages of the reinforcement learning interaction process. 
The \texttt{Buffer} serves as a fundamental data structure for organizing raw data trajectories, while the \texttt{Trainer} and \texttt{Evaluator} act as executors, managing the entire process.
In the following, we will delve into the designs of these core modules. The details of the training and evaluation pipeline can be found in Section 5.

\subsection{Environment}
\label{subsec: env}

The \texttt{Environment} module is responsible for constructing RL environments from static datasets and providing feedback on upcoming actions. Compared to simulation platforms, environments built from public datasets are much lighter and allow for faster training.

Previous work~\cite{DORL-gao2023alleviating} has pointed out that some recommendation datasets are too sparse or lack necessary information (e.g., timestamps, explicit feedback, item categories) to build RL environments.
In EasyRL4Rec, we choose five public datasets suitable for the RL task to construct environments, encompassing diverse recommendation scenarios, such as e-commerce, movies, and short-video recommendations.
The statistics of datasets after preprocessing are summarized in Table~\ref{tab:dataset}.

Consistent with prior research~\cite{huang2020sofa, wang2023rl4rs, zhao2023kuaisim}, we implement RL environments using the APIs of OpenAI Gymnasium~\cite{towers_gymnasium_2023}. The central function of the environment is the \textit{step()} method, which returns observed states and rewards for the current action. Rewards can be defined as clicks, ratings, etc., and can be sourced from static datasets. For offline evaluation, we employ an MF model trained on the test set to predict the rating/reward of vacant user-item pairs.

\begin{table}[h!]
\normalsize
  \centering
  \caption{Datasets currently involved in the EasyRL4Rec. These datasets vary in size and encompass diverse domains.}
  \label{tab:dataset}
  \begin{tabular}{lccccc}
    \toprule
    Dataset & Domain & Usage & \#user & \#item & \#inter.\\
    \midrule
    \multirow{2}*{Coat~\cite{Coat}} & \multirow{2}*{Product} 
        & Train & 290 & 300 & 7.0k \\
      & & Test & 290 & 300 & 4.6k \\
    \hline
    \multirow{2}*{YahooR3~\cite{Yahoo}} & \multirow{2}*{Music} 
    & Train & 15,400 & 1,000 & 311.7k \\
    & & Test & 5,400 & 1,000 & 54.0k \\
    \hline
    \multirow{2}*{MovieLens\footnotemark{}} & \multirow{2}*{Movie} 
    & Train & 6,040 & 3,952 & 800.4k \\
    & & Test & 6,040 & 3,952 & 200.5k \\
    \hline
    \multirow{2}*{KuaiRec~\cite{gao2022kuairec}} & \multirow{2}*{Video} 
    & Train & 7,176 & 10,728 & 12530.8k \\
    & & Test & 1,411 & 3,327 & 4676.6k \\
    \hline
    \multirow{2}*{KuaiRand~\cite{gao2022kuairand}} & \multirow{2}*{Video} 
    & Train & 26,210 & 7,538 & 1141.6k \\
    & & Test & 27,285 & 7,583 & 1186.1k \\
  \bottomrule
\end{tabular}
\end{table}
\footnotetext{\url{https://grouplens.org/datasets/movielens/1m}}

\subsection{Policy}
\label{subsec: policy}

The \texttt{Policy} module applies a RL algorithm to select the optimal action based on the current state. We implement this module by extending RL policies in the Tianshou~\cite{tianshou}, incorporating the following tailored designs for recommendation scenarios.

Firstly, we support both discrete and continuous action-based policies. Given that items in RSs are discrete and better suited for discrete actions, we include a mechanism to convert continuous actions to discrete items, supporting continuous action-based policies.
Secondly, we customize policies by encoding the state via \texttt{StateTracker} since states cannot be directly obtained from environments. These encoded state embeddings are optimized simultaneously with policies.
Thirdly, we introduce a \textit{Remove Recommended Items} option when interacting with environments, addressing the common need for multi-round recommendation. This feature is implemented by adding a mask for logits.

Policies supported by EasyRL4Rec can be categorized as follows:

\begin{itemize}
    \item \textbf{Batch RL}: Learn policies from offline logs collected in advance, also known as OfflineRL. EasyRL4Rec supports classical batch RL algorithms, including BCQ~\cite{DiscreteBCQ-fujimoto2019benchmarking}, CQL~\cite{CQL-kumar2020conservative}, and CRR~\cite{CRR-wang2020critic}.
    \item \textbf{Model-free Off-policy RL}: Learn from trajectory data generated by a different policy. Classic algorithms such as DQN~\cite{DQN-mnih2013playing}, C51~\cite{C51-bellemare2017distributional}, DDPG~\cite{DDPG-lillicrap2015continuous}, and TD3~\cite{TD3-fujimoto2018addressing} have been included in EasyRL4Rec.
    \item \textbf{Model-free On-policy RL}: Learn from trajectory data generated by the current policy being learned. EasyRL4Rec supports algorithms like PG~\cite{REINFORCE-williams1992simple}, A2C~\cite{A2C}, PPO~\cite{PPO-schulman2017proximal}, etc.
\end{itemize}

\subsection{StateTracker}
\label{sec: state_tracker}

Unlike traditional application fields such as games, the state in recommended scenarios cannot be directly obtained from the environment, which requires artificial modeling of the state. The \texttt{StateTracker} module is responsible for modeling and encoding states. 

In most research work~\cite{Self-Supervised-RLinRec-Xinxin, zhang2022multi, CIRS-gao2022cirs, offpolicy-chen2022off}, users' characteristics and interaction history are usually transformed into state representations to capture user preferences and action sequence values at the current moment. This setting will be used for state encoding in this study.
We have implemented five different StateTrackers in EasyRL4Rec, which are all classical models for sequential recommendation: 

\begin{itemize}
    \item \textbf{Average}~\cite{StateRepresentation-liu2020state}. Average concatenates the user embedding and the average pooling result of historical actions as the state representation.
    \item \textbf{GRU}~\cite{GRU4Rec-hidasi2015session}: GRU is a seminal method using RNNs to
model user action sequences for session-based recommendation.
    \item \textbf{Caser}~\cite{Caser-tang2018personalized}: Caser learns sequential patterns using Convolutional Neural Network (CNN) modeling historical actions as an “image” among time and latent dimensions.
    \item \textbf{SASRec}~\cite{SASRec-kang2018self}: SASRec is a self-attention-based sequential model that can balance short-term intent and long-term preference.
    \item \textbf{NextItNet}~\cite{NextItNet-yuan2019simple}: NextItNet is an effective generative model that is capable of learning high-level representation from both short- and long-range item dependencies.
\end{itemize}

\subsection{Collector}

The \texttt{Collector} module serves as a crucial link facilitating interactions between \texttt{Environment} and \texttt{Policy}, responsible for collecting interaction trajectories into \texttt{Buffer}.

\texttt{Collector} plays a pivotal role in both the Training and Evaluation stages. To be specific, at time $t$, the \texttt{Collector} would call \texttt{Policy} to execute an action $a_t$ according to observed information $o_t$. Subsequently, it conveys this action to \texttt{Environment} and gets corresponding rewards $r_t$. 
Considering a complete interaction from time $1$ to time $T$,  the observations, actions, and rewards at each timestamp, denoted as  $\{(o_1, a_1, r_1), ..., (o_T, a_T, r_T)\}$, would be considered as one single trajectory and stored in \texttt{Buffer}. These trajectories would be sampled for policy learning and subsequent updates.

\begin{figure}[t!]
  \centering
  \includegraphics[trim={0 0 0 0}, clip, width=0.45\textwidth]{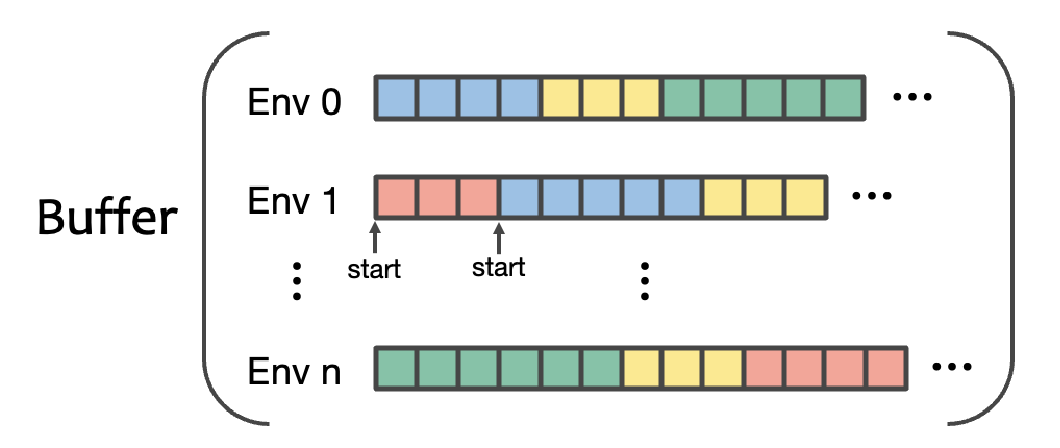}
  \caption{Visualization of data/trajectories stored in \texttt{Buffer}. To support simultaneous interactions in multiple environments, \texttt{Buffer} comprises interaction data from $n$ environments, with different trajectories in each environment represented by distinct colors and the presence of the \textit{start} symbol.}
  \label{fig:data in buffer}
\end{figure}

For effective construction and utilization, EasyRL4Rec supports simultaneous interaction in isolative multiple environments.
As visualization in Figure~\ref{fig:data in buffer}, data in the \texttt{Buffer} are stored in a streaming manner. Data collected at timestamp $t$  would be stored as a block containing all pertinent information (like $(o_t, a_t, r_t)$). These blocks are organized sequentially, and trajectories are differentiated by the presence of the \textit{start} symbol. 

\section{Training and Evaluation Pipeline}

In this section, we will introduce the whole pipeline of training and evaluation applied to RL-based RSs in our library. Following training and evaluation settings exhibit substantial distinctions from conventional recommender systems.

\subsection{Training}
\label{sec: Training}

Different from multi-armed bandit (MAB) algorithms~\cite{UCB-auer2002finite, Egreedy-tokic2010adaptive, ThompsonSampling-chapelle2011empirical, LinUCB-li2010contextual}, which often have simpler problem structure and operational mechanisms, deep RL uses deep neural networks to approximate the optimal policy and/or value functions, requiring training before deployment. 
Typically, training involves an iterative process where the model interacts with an environment to update the network's weights and learn optimal policies through trial and error.

\begin{figure}[t!]
\centering
\includegraphics[trim={0 0 0 0}, clip, width=0.45\textwidth]{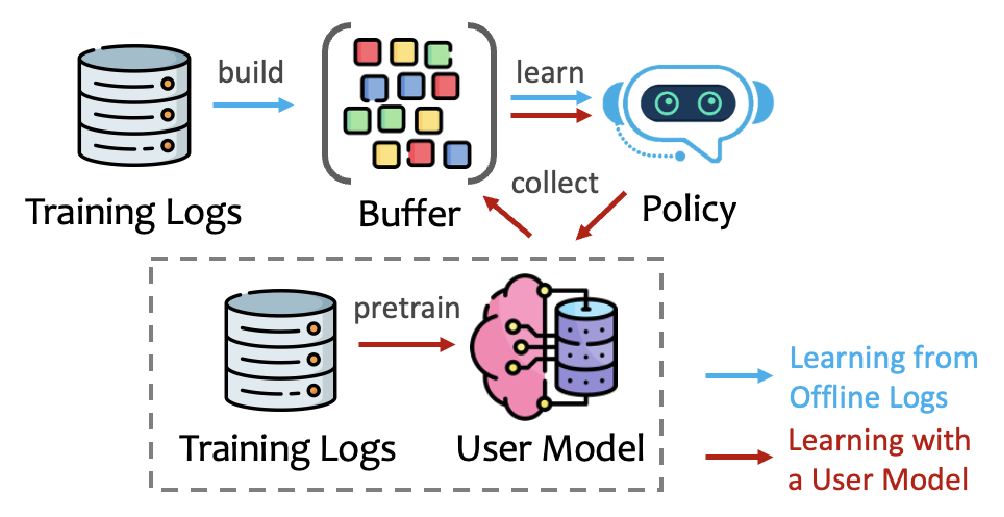}
\caption{Two Training Settings. Blue lines represent the process of learning from offline logs, while red lines represent the process of learning with a user model.}
\label{fig:pipeline-train}
\end{figure}

EasyRL4Rec offers two training settings: (1) Learning from Offline Logs and (2) Learning with a User Model.
In the former setting, the policy directly learns from offline logs, which have been collected in the \texttt{Buffer} in advance. 
As the blue lines shown in Figure~\ref{fig:pipeline-train}, offline logs from the dataset would be split into trajectories and used to build \texttt{Buffer} for the training process.
Then, \texttt{Policy} would iteratively get a batch of trajectories and learn the relationships between actions and outcomes involved.
In EasyRL4Rec, we implement three buffer construction methods: (1) Sequential: logs would be split in chronological order. (2) Convolution: logs would be augmented through convolution. (3) Counterfactual: logs would be randomly shuffled over time. 
This setting is suitable for classic batch RL methods such as BCQ~\cite{BCQ-fujimoto2019off}, CQL~\cite{CQL-kumar2020conservative}, and CRR~\cite{CRR-wang2020critic}.

The Learning with a User Model setting follows a paradigm similar to ChatGPT's RLHF learning~\cite{rlhf-christiano2017deep}. As the red lines shown in Figure~\ref{fig:pipeline-train}, a user model (or reward model) is pre-trained using training data to capture users' preferences. 
Then, a behavior policy would interact with this user model and \texttt{Collector} collects feedback on a series of actions into \texttt{Buffer}. Subsequently, the target policy would learn from trajectories stored in \texttt{Buffer}. The distinction between on-policy and off-policy methods hinges on whether the behavior policy is the same as the target policy. One epoch in the training process can have multiple above loops.

\subsection{Evaluation}

The evaluation process and metrics vary across current studies, making it challenging to compare model performance from different research teams fairly.
In EasyRL4Rec, we execute evaluation in an offline manner, due to less time- and money-consuming than online tests. More importantly, a policy requires thorough evaluation in an offline environment before deployment online.
Due to the sparsity of the dataset and missing interaction data in the test set, we follow the simulation methodology proposed by previous research~\cite{wang2023rl4rs, huang2020sofa}. To be specific, a simulated environment is established to provide missing feedback from users on specific items recommended by the policy. In our library, we adopt a similar approach to ~\cite{huang2020sofa}, leveraging a Matrix Factorization (MF)~\cite{MF1-koren2008factorization, MF2-koren2009matrix} model to predict missing values in the user-item matrix.

To better simulate real online user behavior for evaluation, we introduce a quit mechanism following work~\cite{CIRS-gao2022cirs, quit-xu2022dynamic, DORL-gao2023alleviating}. In this setting, users will interrupt the process of interaction and quit when the termination condition is triggered. The termination condition can be customized according to datasets and requirements of researchers, like considering the mentality of boredom.
Moreover, we offer the option of allowing repeated recommendations, catering to the needs of online applications.

Combining these two mechanisms, EasyRL4Rec offer three modes for evaluation: FreeB, NX\_0\_, NX\_10\_, which are described in detail as follows:

\begin{itemize}
    \item \textbf{FreeB}: allow repeated recommendations, interactions are terminated by quit mechanism.
    \item \textbf{NX\_0\_}: prohibit repeated recommendations, interactions are terminated by quit mechanism.
    \item \textbf{NX\_X\_}: prohibit repeated recommendations, interactions are fixed as X rounds without quit mechanism.
\end{itemize}


Furthermore, EasyRL4Rec provides abundant evaluation metrics that are commonly used in the field of Reinforcement Learning and Recommender Systems, which are summarised in Table~\ref{tab:metrics}. 
To evaluate the long-term effects of RL policies, we introduce Cumulative Reward ($\text{R}_\text{cumu}$) $\sum_t r_t$ to measure the cumulative gain of one interaction trajectory, which is usually applied in the RL scenario. In addition, Interaction Length ($Length$) and Average Reward ($\text{R}_\text{avg}$ measure the length of the interaction trajectory and the single-round reward, respectively.
We also provide commonly used metrics in traditional RSs for evaluating user models or other recommendation models, such as Normalized Discounted Cumulative Gain (NDCG~\cite{ndcg-jarvelin2002cumulated}), HitRate, etc.

\begin{table}[t!]
  \normalsize
  \tabcolsep=7.2pt
  \centering
  \caption{Metrics currently involved in the EasyRL4Rec. Metrics measuring long-term effects are used to evaluate RL policies, while others can be used to evaluate user models.}
  \label{tab:metrics}
  \begin{tabular}{lccc}
    \toprule
    \multicolumn{2}{c}{Scenarios} & Metrics \\
    \midrule
    \makecell[c]{Reinforcement \\ Learning} & \makecell[c]{Long-term \\ effects} & \makecell[c]{Cumulative Reward ($\text{R}_\text{cumu}$), \\ Average Reward ($\text{R}_\text{avg}$), \\ Interaction Length ($Length$)} \\
    \midrule
    \multirow{4}{*}{\makecell[c]{Recommander \\ Systems}} & Prediction & MAE, MSE, RMSE\\
       & Top-\textit{K} & \makecell[c]{Recall, Precision, NDCG, \\ HitRate, MAP, MRR} \\
       & Others & Coverage, Diversity, Novelty\\
    \bottomrule
  \end{tabular}
\end{table}

\section{Application Examples of EasyRL4Rec}

In this section, we present the typical usage example of our library, which includes three stages: Preparation, Training, and Evaluation.
To implement a new algorithm using EasyRL4Rec, researchers can modify our core modules easily. 


\subsection{Initialization \& Preparation}

In this, we must prepare the chosen dataset and train the user model, which will be used to construct environments.
The following code snippet demonstrates the usage of our library to pre-train a user model. Firstly, one must specify the save path and the dataset. After configuring the learning task and initializing the user model, the model can be fitted to the training data. All intermediate results and model parameters will be saved.

\begin{tcblisting}{
    listing engine=minted,
    boxrule=0.1mm,
    colback=blue!5!white,
    colframe=blue!75!black,
    listing only,
    left=5mm,
    minted language=python,
    minted options={
        fontsize=\scriptsize,
        fontfamily=zi4, 
        breaklines, 
        autogobble,
        linenos,
        numbersep=3mm}
}
# 1. Prepare the saved path.
MODEL_SAVE_PATH, logger_path = prepare_dir_log(args)

# 2. Prepare dataset
env, dataset, kwargs_um = get_true_env(args)
dataset_train, dataset_val = prepare_dataset(args, dataset,...)

# 3. Setup user model
task, task_logit_dim, is_ranking = get_task(args.env, args.yfeat)
ensemble_models = setup_user_model(args,...)

# 4. Learn and evaluate model
ensemble_models.fit_data(dataset_train, dataset_val,...)
\end{tcblisting}

\subsection{Policy Training}


As mentioned in section~\ref{sec: Training}, EasyRL4Rec supports two types of training settings. Here, we take the Learning with a User Model setting as an example, with the main code outlined below. We call the \texttt{Policy} module to give a batch of actions according to current states, then obtain feedback from \texttt{Environment}. With trajectories collected in \texttt{Buffer}, we can update the parameters in \texttt{Policy}. 


\begin{tcblisting}{
    listing engine=minted,
    boxrule=0.1mm,
    colback=blue!5!white,
    colframe=blue!75!black,
    listing only,
    left=5mm,
    minted language=python,
    minted options={
        fontsize=\scriptsize,
        fontfamily=zi4, 
        breaklines, 
        autogobble,
        linenos,
        numbersep=3mm}
}
# Collect training data in Buffer
result = self.policy(self.data, self.buffer, ...) # inference
act = to_numpy(result.act)
if self.exploration_noise: # exploration
    act = self.policy.exploration_noise(act, self.data)  
# Obtain feedback from Environment
obs_next, rew, terminated, truncated, info = self.env.step(...) 
...
# Update Policy
losses = self.policy.update(self.batch_size, ...)
\end{tcblisting}

\subsection{Policy Evaluation}

During the evaluation process, we will evaluate the performance of the trained policy, measuring the long-term effects.
As the following code snippet shows, we first reset all parameters in \texttt{Environment} and \texttt{Buffer}. Then, the \texttt{Collector} will call the \texttt{Policy} and \texttt{Environment} module to collect test trajectories, similar to the training process. For collected data, we calculate metrics through callback functions\footnote{A callback is a function provided as an argument to another function, executed after the latter completes its task.}. 

\begin{tcblisting}{
    listing engine=minted,
    boxrule=0.1mm,
    colback=blue!5!white,
    colframe=blue!75!black,
    listing only,
    left=5mm,
    minted language=python,
    minted options={
        fontsize=\scriptsize,
        fontfamily=zi4, 
        breaklines, 
        autogobble,
        linenos,
        numbersep=3mm}
}
# Reset the Environment and Buffer
collector.reset_env()
collector.reset_buffer()
policy.eval()
# Collect test trajectories (call Policy and obtain feedback)
test_result = collector.collect(n_episode) # similar as training
...
# Callback functions to calculate metrics
for callback in self.policy.callbacks:
    callback.on_epoch_end(self.epoch, test_result)
\end{tcblisting}
\section{Experiments}
\label{sec: Benchmark}
In this section, we conduct comprehensive experiments on models based on classic RL policies and some recent work. To ensure a balanced comparison, we separately evaluate the overall performance of model-free RL policies (see Section~\ref{subsec: model-free RL}) and batch RL policies (see Section~\ref{subsec: Batch RL}) under different training conditions. 
We then detail our insights regarding the effectiveness of RL policies in terms of coverage, diversity, and novelty - factors often overlooked in prior research. 
Importantly, we identify the \textit{Preference Overestimation} issue in RL-based RSs, exploring possible reasons.
Further experiments on the influence of each component are available in Section~\ref{sec:appendix-statetrackers} and Section~\ref{sec:appendix-construction}.
All tables and figures can be reproduced by the code available at \url{https://github.com/chongminggao/EasyRL4Rec/tree/main/visual_results}.

\begin{table*}[t!]
\normalsize
\centering
\caption{Performance comparison between model-free RL methods trained with a user model on three datasets. Policies using continuous action are indicated by the symbol (C). The meaning of underlining and bold should be included in the caption. The best results are highlighted in bold, and the second-best results are underlined.}
\label{tab: model-free RL performance}
\tabcolsep=6.5pt
\begin{tabular}{clccccccccc}
\toprule
\multicolumn{2}{c}{\multirow{2}{*}{Method}} & \multicolumn{3}{c}{Coat} & \multicolumn{3}{c}{MovieLens} & \multicolumn{3}{c}{KuaiRec}\\
\cmidrule(lr){3-5}\cmidrule(lr){6-8}\cmidrule(lr){9-11}
\multicolumn{2}{c}{} & $\text{R}_\text{cumu}$ & $\text{R}_\text{avg}$ & Length & $\text{R}_\text{cumu}$ & $\text{R}_\text{avg}$ & Length & $\text{R}_\text{cumu}$ & $\text{R}_\text{avg}$ & Length \\
\midrule
\multirow{4}{*}{Offpolicy} 
 & DQN       &              $54.3476$ &              $2.4687$ &              $22.0196$ &              $21.9550$ &              $2.9400$ &               $7.4624$ &              $12.6543$ &              $0.7935$ &              $15.9480$ \\
 & C51       &              $41.0788$ &              $2.5941$ &              $15.8280$ &              $18.0006$ &              $2.8290$ &               $6.3440$ &              $11.5855$ &              $0.8151$ &              $14.2304$ \\
 & DDPG(C)   &              $16.3348$ &              $2.3277$ &               $7.0200$ &               $9.3706$ &              $3.0329$ &               $3.1152$ &               $9.2155$ &     $\mathbf{1.0192}$ &               $9.0440$ \\
 & TD3(C)    &              $16.3232$ &              $2.3542$ &               $6.9324$ &              $10.1620$ &              $2.9410$ &               $3.4568$ &               $7.8179$ &  \underline{$0.8610$} &               $9.0980$ \\
\midrule
\multirow{8}{*}{Onpolicy}  
 & PG        &  \underline{$79.3392$} &  \underline{$2.6586$} &              $29.8424$ &              $27.9514$ &              $3.2614$ &               $8.5708$ &              $18.8922$ &              $0.6326$ &              $29.8628$ \\
 & A2C       &     $\mathbf{81.7952}$ &     $\mathbf{2.7341}$ &     $\mathbf{29.9164}$ &  \underline{$32.4296$} &              $3.2526$ &               $9.9704$ &     $\mathbf{25.2442}$ &              $0.8437$ &  \underline{$29.9196$} \\
 & PPO       &              $73.0300$ &              $2.5306$ &              $28.8552$ &              $29.2253$ &  \underline{$3.6532$} &               $8.0000$ &              $19.0767$ &              $0.6359$ &     $\mathbf{30.0000}$ \\
 & PG(C)     &              $21.0912$ &              $2.5914$ &               $8.1424$ &              $17.8453$ &              $2.5247$ &               $7.0824$ &              $15.8942$ &              $0.6637$ &              $23.9260$ \\
 & A2C(C)    &              $24.5980$ &              $2.5137$ &               $9.7932$ &              $26.5039$ &              $3.3884$ &               $7.8172$ &              $18.2968$ &              $0.6732$ &              $27.2416$ \\
 & PPO(C)    &              $53.2212$ &              $2.5962$ &              $20.5100$ &              $29.4684$ &     $\mathbf{3.8737}$ &               $7.6016$ &              $18.6928$ &              $0.6730$ &              $27.6780$ \\
 & DORL      &              $76.9936$ &              $2.6025$ &              $29.5816$ &     $\mathbf{45.7708}$ &              $2.6401$ &     $\mathbf{17.3440}$ &  \underline{$22.5246$} &              $0.7533$ &              $29.9016$ \\
 & IntRD &              $77.4292$ &              $2.5926$ &  \underline{$29.8660$} &              $25.9168$ &              $2.2783$ &  \underline{$11.3676$} &              $20.9392$ &              $0.7748$ &              $27.0216$ \\

\bottomrule
\end{tabular}
\end{table*}

\subsection{Experimental Settings}
\label{subsec: experimental settings}

\subsubsection{\textbf{Datasets}}
We conduct experiments on Coat\footnote{\url{https://www.cs.cornell.edu/~schnabts/mnar/}}, MovieLens\footnote{\url{https://grouplens.org/datasets/movielens/1m/}}, and KuaiRec\footnote{\url{https://kuairec.com/}}, three representative datasets with various scales and domains. Statistics of three datasets have been shown in Table~\ref{tab:dataset}. Details of data Preprocessing and environment building are presented in Section~\ref{subsec: env}.

\subsubsection{\textbf{Models implementation}}
Currently, EasyRL4Rec supports more than fifteen classic RL policies as base models and implements several recent work~\cite{Self-Supervised-RLinRec-Xinxin, DORL-gao2023alleviating} in RL-based RSs.
Experiments are conducted on some representative algorithms, covering batch RL policies and model-free RL policies.
Reproducing more models in previous work is considered future work.

\subsubsection{\textbf{Experimental Details}}
In the training stage, all policies are trained with 100 epochs, with the default learning rate 1e$^{-3}$. We apply the first training paradigm, i.e. learning from offline logs, to train batch RL algorithms. When training model-free RL algorithms, we pre-train a user model (DeepFM~\cite{guo2017deepfm}) to guide the online learning of RL policies.
For the evaluation process, after each epoch, the policy undergoes evaluation using 100 episodes (i.e., interaction trajectories), and the maximum recommended sequence length is limited to 30.
We set the random seed to 2023 for consistency and reproducibility. The code for replicating all experimental results is available in our library.

\subsection{Performance of Model-free RL}
\label{subsec: model-free RL}

We evaluate the performance of representative model-free RL algorithms, which can be divided into two groups:

\begin{itemize}[leftmargin=*]
    \item \textbf{Off-policy}: Off-policy methods learn from data produced by a policy different from the one currently being optimized. For our experiments, we choose Q-learning based \textbf{DQN}~\cite{DQN-mnih2013playing}, \textbf{C51}~\cite{C51-bellemare2017distributional}, and continuous control based methods like \textbf{DDPG}~\cite{DDPG-lillicrap2015continuous} and \textbf{TD3}~\cite{TD3-fujimoto2018addressing}.
    \item \textbf{On-policy}: Contrarily, on-policy methods utilize data generated by their current policy. This category includes policy gradient-based \textbf{PG}~\cite{REINFORCE-williams1992simple}, and actor-critic-based \textbf{A2C}~\cite{A2C} and \textbf{PPO}~\cite{PPO-schulman2017proximal}.
    \item \textbf{Others}: \textbf{DORL}~\cite{DORL-gao2023alleviating} is a debiased model based on A2C policy to alleviate the Matthew effect. \textbf{Intrinsic} modifies the reward model by incorporating an intrinsic reward, which fosters exploration to enhance item coverage and diversity.
\end{itemize}

While items in Recommender Systems (RSs) are inherently discrete and thus more compatible with discrete action-based policies, EasyRL4Rec incorporates a feature that enables the transformation of continuous actions into discrete items, supporting continuous action-based policies.
The overall performance of model-free RL algorithms is presented in Table~\ref{tab: model-free RL performance}, which reports the mean values of all metrics during the last 25\% of the training epochs.
From the experimental results, we mainly have the following observations.

Firstly, discrete action-based methods perform much better than continuous action-based methods (marked by (C)). 
Discrete methods such as DQN and C51 in the off-policy category, and PG and PPO in the on-policy category, achieve higher $\text{R}_\text{cumu}$ and longer $Length$ across all three datasets.
Yet continuous methods like DDPG(C), TD3(C), PG(C), and A2C(C) show fewer Lengths, which might be due to less efficient exploration strategies or poor fit between the continuous action representation and the discrete item space in RSs.
This trend suggests that discrete action spaces, which are more aligned with the discrete items in recommender systems, are more effective for model-free RL algorithms.

\begin{figure}[t!]
\centering
\includegraphics[trim={0 0 0 0}, clip, width=0.48\textwidth]{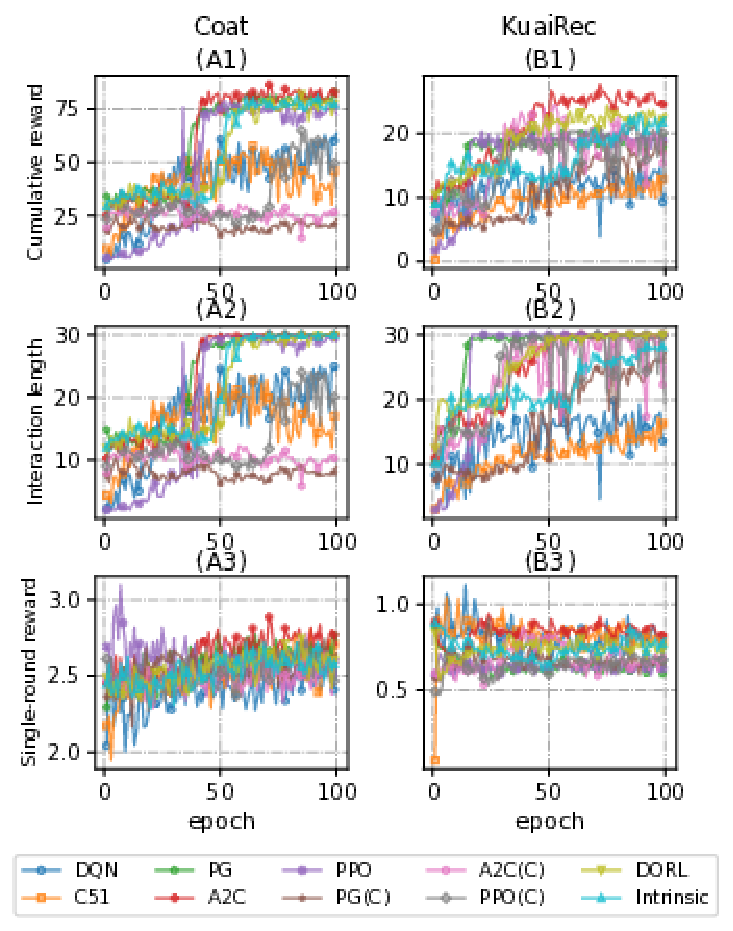}
\caption{Variation curves of 1) cumulative reward, 2) interaction length, and 3) single-round reward during training on the Coat and KuaiRec datasets.}
\label{fig:training curve}
\end{figure}

Secondly, on-policy methods achieve better results than off-policy methods within the same action type. For example, focusing on discrete action strategies on the KuaiRec dataset, the on-policy method PG demonstrates a better performance with an $\text{R}_\text{cumu}$ of compared to the off-policy method DQN. This trend is further supported by Figure~\ref{fig:training curve}, which shows the variation curves of 1) cumulative reward, 2) interaction length, and 3) single-round reward during training on the Coat and KuaiRec datasets.
This could indicate that on-policy methods, which keeps target-policy align with behavior-policy, are more adept at handling the exploration-exploitation tradeoff in these contexts.

Thirdly, methods designed for RSs (DORL, Intrinsic) achieve competing results to the best method. On the Coat and KuaiRec datasets, DORL and Intrinsic achieve slightly worse results than A2C, while on MovieLens, DORL achieves superior performance compared to other models.

\subsection{Performance of Batch RL}
\label{subsec: Batch RL}

We conducted experiments on models based on Batch Reinforcement Learning (Batch RL) using the following representative algorithms:

\begin{itemize}[leftmargin=*]
    \item \textbf{BCQ}~\cite{DiscreteBCQ-fujimoto2019benchmarking}, short for Batch-Constrained deep Q-learning, utilizes high-confidence data to update the policy.
    \item \textbf{CQL}~\cite{CQL-kumar2020conservative}, or Conservative Q-Learning, incorporates a Q-value regularizer over an actor-critic policy.
    \item \textbf{CRR}~\cite{CRR-wang2020critic}, or Critic Regularized Regression, trains the policy to avoid out-of-distribution (OOD) actions.
    \item \textbf{SQN}~\cite{Self-Supervised-RLinRec-Xinxin}, or Self-Supervised Q-learning, employs self-supervised learning (SSL) to enhance Reinforcement Learning-based Recommender Systems (RL RSs). It comprises two output layers (heads): one for the cross-entropy loss and the other for RL.
\end{itemize}

\begin{table}[t!]
\small
\centering
\caption{Performance comparison between batch RL methods trained from offline logs. The best results are highlighted in bold, and the second-best results are underlined.}
\label{tab: batch RL performance}
\tabcolsep=3.5pt
\begin{tabular}{lccccccccc}
\toprule
\multirow{2}{*}{Method} & \multicolumn{3}{c}{Coat} & \multicolumn{3}{c}{MovieLens} & \multicolumn{3}{c}{KuaiRec} \\
\cmidrule(lr){2-4}\cmidrule(lr){5-7}\cmidrule(lr){8-10}
 & $\text{R}_\text{cumu}$ & $\text{R}_\text{avg}$ & Len & $\text{R}_\text{cumu}$ & $\text{R}_\text{avg}$ & Len & $\text{R}_\text{cumu}$ & $\text{R}_\text{avg}$ & Len \\
\midrule
BCQ &                 $9.56$ &       $\mathbf{2.36}$ &               $4.06$ &    $9.54$ &    \underline{$3.81$} &              $2.51$ &                 $4.34$ &                $0.83$ &              $5.24$ \\
CQL &                $23.35$ &                $2.27$ &              $10.28$ &       \underline{$9.89$} &       $\mathbf{3.85}$ &              $2.57$ &                 $5.28$ &       $\mathbf{1.03}$ &              $5.12$ \\
CRR &       $\mathbf{28.96}$ &                $2.23$ &     $\mathbf{13.00}$ &                $\mathbf{10.13}$ &                $2.95$ &     $\mathbf{3.44}$ &        $\mathbf{8.57}$ &    \underline{$0.87$} &     $\mathbf{9.83}$ \\
SQN &    \underline{$26.60$} &    \underline{$2.31$} &  \underline{$11.51$} &                 $9.45$ &                $2.78$ &  \underline{$3.39$} &     \underline{$6.82$} &                $0.77$ &  \underline{$8.92$} \\
\bottomrule
\end{tabular}
\end{table}

Table~\ref{tab: batch RL performance} shows the performance of different batch RL algorithms trained from offline logs.
As shown, CRR outperforms all models across all three datasets, while BCQ yields the lowest performance. 
Regarding the issue of Preference Overestimation (detailed in Section~\ref{subsec: Overestimation}), both CQL and CRR outperform BCQ due to their more conservative strategies.
Furthermore, it is worth noting that due to the limited availability of offline logs, batch RL methods achieve lower rewards compared to model-free RL baselines that are trained online with a user model.

\subsection{Coverage, Diversity \& Novelty}

We evaluate the effectiveness of these models in terms of coverage, diversity, and novelty—factors often overlooked in prior research. From Table~\ref{tab: cov} we can observe the performance of on-policy and off-policy methods with respect to coverage, diversity, and novelty.

\begin{table}[t!]
\normalsize
\centering
\tabcolsep=7.5pt
\caption{Performance of Coverage, Diversity \& Novelty on the KuaiRec Dataset. The best results are highlighted in bold, and the second-best results are underlined.}
\label{tab: cov}
\begin{tabular}{clccccccccc}
\toprule
\multicolumn{2}{c}{Method} & Coverage & Diversity & Novelty  \\
\midrule
\multirow{4}{*}{Offpolicy} 
 & DQN       &  \underline{$0.0505$} &     $\mathbf{0.8981}$ &              $2.6370$ \\
 & C51       &     $\mathbf{0.1027}$ &              $0.8797$ &              $2.5135$ \\
 & DDPG(C)   &              $0.0477$ &              $0.8209$ &              $2.3907$ \\
 & TD3(C)    &              $0.0438$ &              $0.8286$ &              $2.2612$ \\
\midrule
\multirow{8}{*}{Onpolicy}
 & PG        &              $0.0015$ &              $0.8273$ &              $2.9794$ \\
 & A2C       &              $0.0020$ &              $0.8291$ &              $2.3907$ \\
 & PPO       &              $0.0015$ &              $0.8276$ &              $2.5408$ \\
 & PG(C)     &              $0.0071$ &              $0.8386$ &              $3.0606$ \\
 & A2C(C)    &              $0.0098$ &              $0.8419$ &  \underline{$3.4543$} \\
 & PPO(C)    &              $0.0141$ &              $0.8677$ &     $\mathbf{3.5830}$ \\
 & DORL      &              $0.0021$ &              $0.8307$ &              $3.1160$ \\
 & Intrinsic &              $0.0274$ &  \underline{$0.8950$} &              $2.7575$ \\
\bottomrule
\end{tabular}

\end{table}

For coverage, which measures the proportion of the state-action space that the policy explores, off-policy methods exhibit higher coverage than the on-policy methods. This observation indicates that on-policy methods are generally more conservative in exploration. In terms of diversity, which quantifies the variety of the actions taken by the policy, all methods perform similarly. Among all methods, DQN stands out with a diversity score of 0.8981, suggesting it is capable of producing a wide range of actions. It's worth noting that Intrinsic promotes better coverage and diversity than its on-policy counterparts due to its reward structure.
Novelty measures the tendency of the policy to recommend less popular or less known items. As we can see, on-policy methods perform better than off-policy methods, which indicates that on-policy methods pay more attention to the user's niche interests.


\subsection{Preference Overestimation Issue}
\label{subsec: Overestimation}
In this section, we discuss our observations on the \textit{Preference Overestimation} issue that occurs in RL-based RSs. Much like the challenge of value overestimation in offline RL, this particular issue can result in an exaggerated estimation of user preferences for items that rarely appear in the training logs. To explore this problem, we evaluated the performance of the A2C algorithm using various user models, each trained with a different number of negative samples.

\begin{figure}[t!]
\centering
\includegraphics[trim={0 0 0 0}, clip, width=0.48\textwidth]{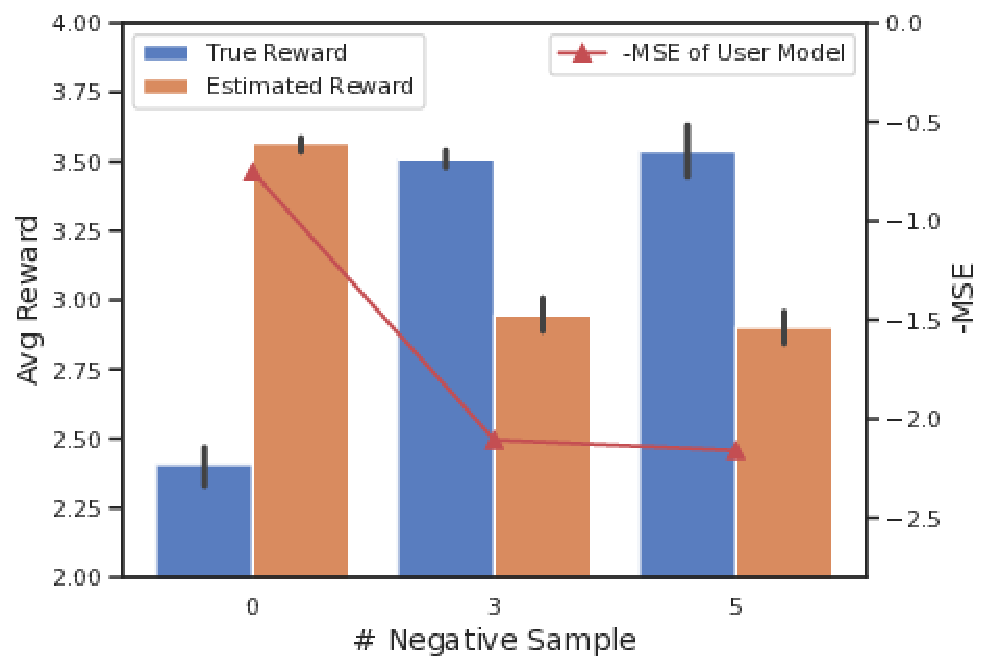}
\caption{Demonstration of preference overestimation issue on MovieLens-1M, with red lines representing -MSE of User Model, while orange bars and blue bars representing estimated reward and true reward respectively.}
\label{fig:value overestimation}
\end{figure}

From Figure~\ref{fig:value overestimation} we can observe that user models trained with \textbf{fewer} negative samples tend to predict more accurately, as indicated by the higher negative mean squared error (MSE). However, choosing a user model with seemingly superior predictive performance (for instance, one trained with 0 negative samples) leads to a paradox. Despite the model predicting higher rewards, as shown by the orange bars, the actual average rewards received, represented by the blue bars, are disappointingly low. This stark contrast brings to light the issue of preference overestimation.

Though this issue could be alleviated by some conservative algorithms, our findings suggest that increasing the number of negative samples could be a viable strategy. This approach remains further explored in future research.




\subsection{Impact of StateTrackers}
\label{sec:appendix-statetrackers}

StateTracker is used to generate representations of the current state as input to \texttt{Policy} module. The details of different StateTrackers can be found in Section~\ref{sec: state_tracker}.
We conduct experiments to investigate the impact of the choice of StateTrackers, encompassing today’s popular methods in sequential modeling. 

We keep the same experimental setting as Section~\ref{sec: Benchmark} and choose A2C~\cite{A2C} as the algorithm. From Table~\ref{tab:state_trackers} we can observe that, there is no significant difference between all StateTrackers, with GRU achieving the best $\text{R}_\text{cumu}$. It indicates that the choice of StateTracker has a minimal impact on the model performance.


\begin{table}[H]
  \normalsize
  \tabcolsep=7.2pt
  \centering
  \caption{Performance comparison between StateTrackers on KuaiRec. The best results are highlighted in bold, and the second-best results are underlined.}
  \label{tab:state_trackers}
  \vspace{-2mm}
  \begin{tabular}{lcccccc}
    \toprule
    StateTrackers & $\text{R}_\text{cumu}$ & $\text{R}_\text{avg}$ & Length \\
    \midrule
    GRU~\cite{GRU4Rec-hidasi2015session} & $\mathbf{19.1353}$ & $\mathbf{0.6385}$ & $29.9688$ \\
    Caser~\cite{Caser-tang2018personalized} & $18.6679$ & $0.6226$ & $29.9824$ \\
    SASRec~\cite{SASRec-kang2018self} & $18.5527$ & $0.6184$ & \underline{$30.0000$} \\
    Average~\cite{StateRepresentation-liu2020state}& \underline{$18.8922$} & \underline{$0.6326$} & $29.8628$ \\
    NextItNet~\cite{NextItNet-yuan2019simple} & $18.7167$ & $0.6239$ & $\mathbf{30.0000}$ \\
    \bottomrule
  \end{tabular}
\end{table}


\subsection{Impact of Construction Methods}
\label{sec:appendix-construction}

As mentioned in section~\ref{sec: Training}, 
EasyRL4Rec offers three different buffer construction methods: (1) Sequential: logs are split in chronological order. (2) Convolution: logs are augmented through convolution. (3) Counterfactual: logs are randomly shuffled over time. 

We conduct experiments with CRR~\cite{CRR-wang2020critic} policy on the KuaiRec dataset, with results present in Table~\ref{tab:construction_methods}.
We can observe no significant difference across the Sequential, Convolution, and Counterfactual constructions. The slight variations in performance metrics suggest that all three constructions provide comparable effectiveness in their contributions to the model's ability to predict or recommend.

\begin{table}[H]
  \normalsize
  \tabcolsep=7.2pt
  \centering
  \caption{Performance comparison between different construction methods on KuaiRec. The best results are highlighted in bold, and the second-best results are underlined.}
  \label{tab:construction_methods}
  \begin{tabular}{lcccccc}
    \toprule
    Construction & $\text{R}_\text{cumu}$ & $\text{R}_\text{avg}$ & Length \\
    \midrule
    Sequential & \underline{$8.5656$} & $\mathbf{0.8713}$ & $9.8332$ \\
    Convolution & $8.5639$ & \underline{$0.8704$} & \underline{$9.8420$} \\
    Counterfactual & $\mathbf{8.5740}$ & $0.8701$ & $\mathbf{9.8516}$ \\
    \bottomrule
  \end{tabular}
\end{table}

\section{Conclusion \& Future Work}

In this work, we introduce EasyRL4Rec, an easy-to-use code library specifically crafted for RL-based RSs, simplifying the development and evaluation of RL models. 
EasyRL4Rec constructs lightweight RL environments based on five public datasets.
It offers a unified training and evaluation pipeline, evaluating models from the perspective of long-term benefits.
Moreover, EasyRL4Rec facilitates customizable state modeling and action representation, addressing challenges in applying RL to recommender systems.
Through comprehensive experiments, we compare the performance of existing models and share our findings from various perspectives.




Currently, EasyRL4Rec supports lots of classic RL algorithms, but few recent studies in RL-base RSs.
In the future, we plan to extend EasyRL4Rec to include more existing models, more datasets, and more utils like parameters-tuning for easy usage.
EasyRL4Rec is expected to facilitate future research in the field of RL-based RSs.





\newpage
\onecolumn
\begin{multicols}{2}
\bibliographystyle{ACM-Reference-Format}
\bibliography{bibliography}
\end{multicols}

\end{sloppypar}
\end{document}